\documentclass[11pt,twoside]{article}
\usepackage{asp2010}
\usepackage{graphicx}
\usepackage{wrapfig}
\usepackage{rotating}
\usepackage[rightcaption]{sidecap}

\resetcounters

\begin{document}

\title{Synthetic spectra for O and B type subdwarf stars}
\author{Peter Nemeth$^{1}$, Roy {\O}stensen$^{1}$, Pier-Emmanuel
Tremblay$^2$ and Ivan Hubeny$^3$}
\affil{
$^1$ Instituut voor Sterrenkunde, KU Leuven, Celestijnenlaan 200D,
3001 Leuven, Belgium; \texttt{pnemeth1981@gmail.com}\\
$^2$ Zentrum f\"ur Astronomie der Universit\"at Heidelberg,
Landessternwarte, K\"onigstuhl 12, 69117 Heidelberg, Germany\\
$^3$ Steward Observatory, University of Arizona, 933 North Cherry Avenue,
85721 Tucson, AZ, USA\\}

\begin{abstract}
We present a grid of optical (3200--7200 \AA) synthetic
spectra calculated with {\sc Tlusty/Synspec}.
The new NLTE model atmospheres include the most recent hydrogen Stark
broadening profiles; were calculated in opacity sampling and limited to
pure H/He composition. 
The grid covers the observed parameter space of (He-)sdB and (He-)sdO stars,
therefore it is suitable for the homogeneous
spectral analyses of such evolved stars.
\end{abstract}

\section{Introduction}
Hot subdwarfs are important objects to understand the late stages of
stellar evolution.
These He burning stars connect the red giant branch to the white dwarf
sequence and suffer a strong mass-loss before they settle
on the extreme horizontal branch for about a 100 Myr.
This way hot subdwarfs also give an insight into the
horizontal branch morphology.
The conditions and reasons of the extreme
mass-loss are not easy to decipher, in
particular for single stars.
Current formation theories can explain the observed subdwarf
populations, but
extensive spectral libraries, that include all types of hot subdwarfs,
would be welcome to separate the observed populations
and find the relative contributions of the various formation channels.

Line blanketing and NLTE effects are strong and must be considered
in subdwarf spectroscopy
while the static plane-parallel atmosphere approximations are still valid.
The NLTE model atmosphere code {\sc Tlusty} is long used in subdwarf
atmospheric analyses,
libraries of synthetic
spectra have been calculated and fitted
successfully for numerous subdwarfs.
However, recent
improvements in the theoretical models and the lack of a homogeneous,
publicly available
reference library motivated us to calculate a new grid.
Here we present a comprehensive spectral library, which is based on the
observed distribution of hot subdwarf stars and
suitable for their homogeneous analyses.

\section{Model calculation}
We used {\sc Tlusty 204} and {\sc Synspec 49}
(\citealt{hubeny95}; \citealt{lanz95}) in our work to calculate
plane--parallel, horizontally homogeneous, static stellar atmospheres in
radiative and hydrostatic equilibrium.
Even though metal line blanketing is not present in H/He models we used
opacity sampling for a better distribution of frequency points. 
The most detailed model atoms of H and He were included
from the BSTAR database \citep{lanz07} giving altogether 61 explicit
energy \mbox{(super-)levels} and 330 radiative transitions.

The grid was calculated in about 1400 CPU hours and
consists of 11\,396 grid points to which we were able to
converge 10\,887 spectra (95.5\%) in the current version.
The grid is complete between $\log g = 5.0$ and $6.1$ (cgs), most of the
slow converging models are 
either at low $T_{\rm eff}$ -- high $\log g$ (the extremely low-mass white
dwarf regime) or at high He abundance -- high
$\log g$, where 
no stars were observed in a sample of
$\sim$170 hot subdwarfs selected from the {\sc GALEX} survey
\citep{nemeth12}.
However, that analysis was done with
{\sc Tlusty} as well, which raises the concern of non-detectability of such
stars.
The atmospheres with a high helium abundance and low effective temperature    
are convectively unstable and a He convection zone is difficult to model.     

The radiation fields, and therefore the 
spectral energy distributions, were calculated at 16\,000--18\,000
frequency points in opacity sampling.
We used 70 depth points to discretize the atmospheric structure between the
Rosseland mean opacities $\tau_{\rm Ross}=10^{-7}$ and $\tau_{\rm
Ross}=100$. 
The microturbulent velocity was held zero and all models were
\mbox{converged to a}
relative change of 0.1\% of the structural parameters.
Synthetic spectra
were then calculated in the range of 3200--7200 \AA\ 
with 0.1 \AA\ step size, ensuring 
that both the continuum and the line profiles are sampled well.
Because the abundances were limited to H/He our models do not address the
Balmer line problem. 
Due to this shortcoming the derived atmospheric parameters are not expected
to be consistent with the more accurate photometric or asteroseismic 
constraints.

\begin{sidewaysfigure}
{\includegraphics[angle=0,width=\textheight,clip=]{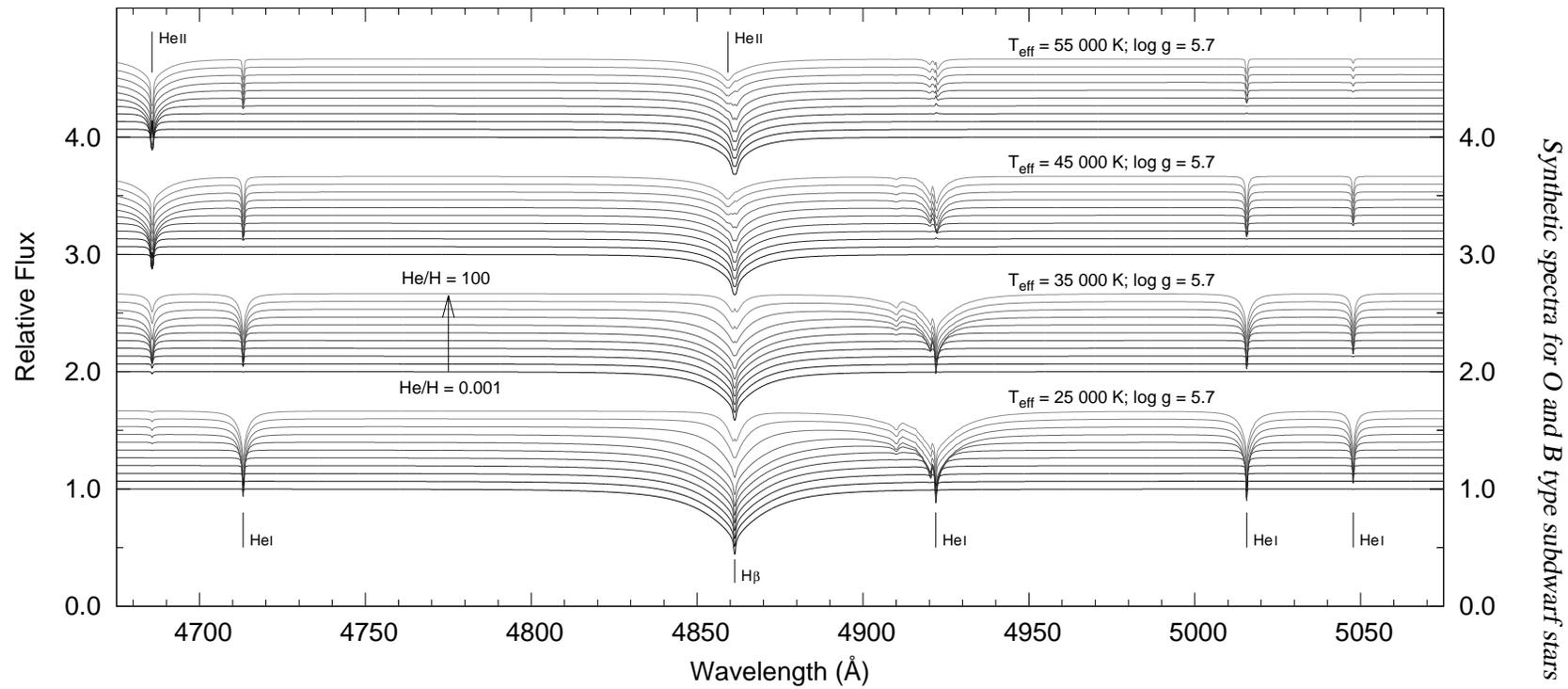}}\\
\caption{  {\small 
Spectral variations along increasing temperature and He abundances from
bottom
to top at $\log g=5.7$ (cgs). } \label{Fig:1}}
\end{sidewaysfigure}

A sample of synthetic spectra near the H$\beta$ line are shown in Figure
\ref{Fig:1}.
Effective temperature increases from bottom to top in the figure as well as
the He
abundance in each sequence. The shift in ionization balance is marked 
by the gradual decrease of the He\,I/He\,II line strength ratio
towards higher temperatures. 
Also remarkable is the weakening of the H$\beta$ line at higher
temperatures and He abundances, while in the most extreme cases the He\,II
Pickering series dominates the spectra.

\section{Line profiles}

\begin{SCfigure*}
\centering
\includegraphics[angle=0,width=0.6\textwidth]{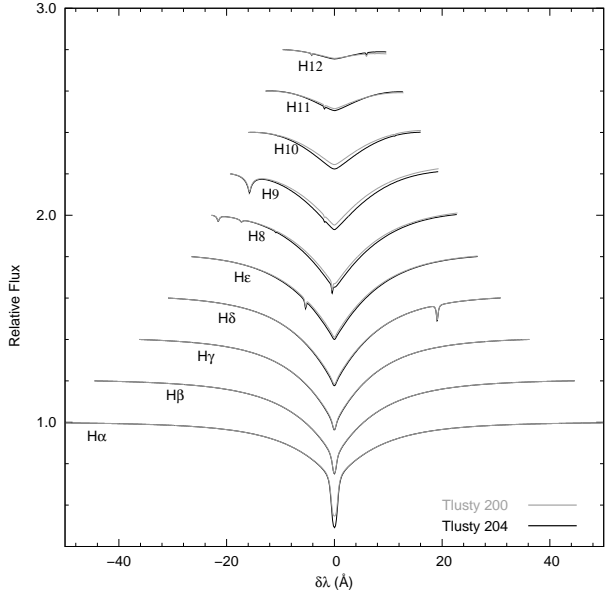}
\caption{  {\small
Hydrogen line profiles of a typical sdB star with $T_{\rm eff}=32\,000 
{\rm K,\ } \log g=5.7\ {\rm (cgs)\ and\ }  {\rm He/H}=0.013$ model. Lines
were shifted for clarity and normalized at the short wavelength end.
\vspace{2em}
}\label{Fig:2} }
\end{SCfigure*}

The new version of {\sc Tlusty} makes it possible to
include detailed H line profiles
directly in the atmosphere structure calculations.
We added the Stark broadening tables for the first ten lines of the Lyman
and Balmer series of H from \citet{tremblay09}. 
A systematic line profile difference towards the high members of the
Balmer series has
been observed in white dwarfs and it is also notable for subdwarfs, in particular
for the H$\epsilon$ -- H${10}$ lines in Figure \ref{Fig:2}. 
The differences in the line profiles are relatively small.
However, because the surface 
gravity is constrained by the upper Balmer series,
even such small differences can be responsible for about half of the
discrepancy observed between
spectroscopic and photometric surface gravities, also known as the $\log g$
problem \citep{rauch12}.
We expect that the new profiles will
increase the internal 
consistency of temperature and gravity
determinations from various Balmer lines.
Surprisingly, we found a deeper H$\alpha$ line core.
Although the new line profiles modify the 
atmospheric structure and the level populations, this change was unexpected
and needs further investigations.

Beyond the H line profiles our models include detailed profiles for the
overlapping forbidden components of four He\,I lines 
($\lambda$4026, $\lambda$4388, $\lambda$4471,
and $\lambda$4922 \AA) as well as detailed profiles for the He\,II Balmer, 
Paschen and Pfund series.
The structure of the $\lambda$4922 He\,I line can be seen 
in Figure \ref{Fig:1}.

\section{The library}

\begin{table*}
\begin{center}
\vspace{-3mm}
\parbox{0.85\textwidth}{\caption{{\small Boundaries and the resolution of the spectral
library.\label{Tab:1}}}}\\
\vspace{1mm}
\begin{tabular}{l|c|c|c} 
     \hline\hline
     Parameter            &Lower limit    & Step size    & Higher limit\\
     \hline
     $T_{\rm eff}$ (K)      &$20\,000$ & $+1000$  &$ 56\,000$ \\
     $\log g     $ (cgs)  &$5.0   $  & $+0.1 $  &$ 6.3    $ \\
     $n{\rm He}/n{\rm H}$ &$0.0005$  & $\times2$&$ 100    $ \\
     \hline\hline
\end{tabular}
\end{center}
\end{table*}

The parameters of our grid are listed in Table \ref{Tab:1}.
The library consists of 
synthetic spectra and continuum flux in {\sc ascii} format
and can be downloaded from:  
\begin{center}
\url{http://www.ster.kuleuven.be/~petern/work/sd_grid/}
\end{center}

\end{document}